\documentclass[review]{elsarticle}
\usepackage{url,hyperref,microtype,subcaption}
\usepackage{hyperref}
\usepackage{amsmath}
\journal{Chaos, Solitons \& Fractals}

\begin{document}

\begin{frontmatter}

\title{Control of coherence resonance in multiplex neural networks}

\author{Maria Masoliver\,$^{1}$ Cristina Masoller\,$^{1}$ and Anna Zakharova\,$^{2}$}
\address{$^{1}$Departament de Fisica, Universitat Politecnica de Catalunya, Rambla Sant Nebridi 22, 08222 Terrassa, Barcelona, Spain\\
$^{2}$ Institut f\"{u}r Theoretische Physik, Technische Universit\"{a}t Berlin, Hardenbergstr. 36, 10623 Berlin, Germany }

\begin{abstract}
We study the dynamics of two neuronal populations weakly and mutually coupled in a multiplexed ring configuration. We simulate the neuronal activity with the stochastic FitzHugh-Nagumo (FHN) model. The two neuronal populations perceive different levels of noise: one population exhibits spiking activity induced by supra-threshold noise (layer 1), while the other population is silent in the absence of inter-layer coupling because its own level of noise is sub-threshold (layer 2). We find that, for appropriate  levels of noise in layer 1, weak inter-layer coupling can induce coherence resonance (CR), anti-coherence resonance (ACR) and inverse stochastic resonance (ISR) in layer 2. We also find that a small number of randomly distributed inter-layer links are sufficient to induce these phenomena in layer 2. Our results hold for small and large neuronal populations.
\end{abstract}


\end{frontmatter}

\section{Introduction}
A fundamental challenge of complexity science is to understand synchronization and emergent phenomena in complex systems represented by sets of excitable units coupled with different topologies. Multilayer networks are receiving increasing attention because they represent many real-world systems \cite{multi1,multi2,LEY18}.
Multilayer networks are composed of interconnected layers, where each layer is formed by a set of $N$ units or nodes, whose interactions are represented by links. In the case when the inter-layer interactions are only vertical, e.g., node $i$ in layer 1 is linked to node $i$ in layer 2, the network is called multiplex. Multiplex networks represent, therefore, a special class of multilayer networks where the layers contain the same number of nodes
and the inter-layer links are allowed only for replica nodes,
i.e., there is a one-to-one correspondence between the nodes in different layers.
In this work, we focus on a two-layer multiplex network where each layer is formed by $N$ neurons coupled in a ring configuration (see Fig.~\ref{fig:esquema_two}).

We investigate the phenomenon of coherence resonance that corresponds to the state of the network characterized by high temporal regularity of noise-induced oscillations achieved for an intermediate optimal noise intensity~\cite{GAN93,PIK97,MAS17,SEM18}. This phenomenon is an example of the constructive role of noise in excitable dynamical systems~\cite{review_2004}. 
Coherence resonance has been reported not only in excitable \cite{JAN04,AUS09,HIZ06,HIZ08b, PIS19}, but also in non-excitable systems \cite{USH05,ZAK10a,ZAK11,ZAK13,GEF14,SEM15}.  In complex networks of FitzHugh-Nagumo units, it has been investigated in one-layer \cite{MAS17} and two-layer \cite{SEM18} networks. Further topologies include local, nonlocal, global coupling, lattice networks as well as more complex structures such as random or small-world networks \cite{WAN00,KWO02,SUN08,YIL16,MAS17,AND18}.  

One of the most relevant and at the same time challenging questions is related to the control of coherence resonance. A well-studied mechanism of coherence resonance control is based on time delay. For example, the control of coherence resonance in a one-layer
network of delay-coupled FitzHugh-Nagumo neurons has been investigated in \cite{MAS17}. Moreover, time-delayed feedback control has been applied to a special type of coherence resonance called coherence-resonance chimera occurring in a
ring of nonlocally coupled excitable FitzHugh-Nagumo
systems \cite{SEM16,ZAK17,ZAK17a}. 

Multilayer networks offer new possibilities of control via the interplay
between dynamics and multiplexing. The advantage of this method is that it allows regulating the dynamics of one layer by adjusting the parameters of the other layer \cite{GHO16,GHO18}. Recently, the so-called weak multiplexing control has been reported and applied to coherence resonance \cite{SEM18} and chimera states \cite{ZAK20,MIK18}. The distictive feature and the advantage of this control scheme is the possibility of achieving the desired state in a certain layer without manipulating its parameters and in the presence of weak coupling between the layers (i.e., the coupling between the layers is smaller than that inside the layers). While the time-delayed feedback control of coherence resonance has been well-understood, the multiplexing control has been much less investigated.

The aim of this work is to study coherence resonance in a two-layer network of FitzHugh-Nagumo neurons \cite{FHN1,FHN2} with weak inter-layer coupling. In particular, we focus on the case of unequally noisy layers: a noisy layer (layer 1) that displays spiking activity induced by supra-threshold noise, is multiplexed with a ``silent'' layer (layer 2), which has subthreshold noise, and whose spiking activity is induced by weak coupling to layer 1. Recently, the possibility of inducing coherence resonance in the silent layer has been shown~\cite{SEM18}. Here we analyze the role of the system size and the impact of removing inter-layer links. We find that not only coherence resonance, but also, anti-coherence resonance (ACR) and inverse stochastic resonance (ISR) can be induced in layer 2. ACR is characterized by high temporal irregularity of noise-induced oscillations~\cite{anticoh} and ISR is characterized by noise suppression of oscillations (the average spiking rate of a neuron exhibits a minimum with respect to noise)~\cite{isr,isr2}. We also find that a small number of randomly distributed inter-layer links can be sufficient to induce these phenomena in layer 2. 

This paper is organized as follows: Sec.~\ref{sec:model} presents the model; sec.~\ref{sec:methods} presents the measures used to quantify the regularity of the neuronal spiking activity, sec.~\ref{sec:results} presents the results and sec.~\ref{sec:conclusions} summarizes our conclusions.

\section{Model} \label{sec:model}

\begin{figure}[t]
\centering
  \includegraphics[width=0.6\columnwidth]{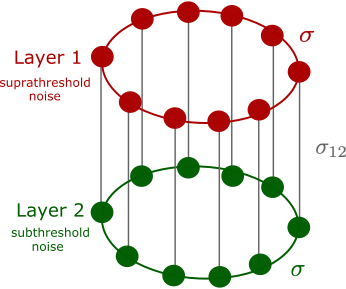}
  \caption{Schematic representation of the network under study: a multiplex neural network consisting of two layers coupled through the inter-layer coupling $\sigma_{12}$. Nodes within each layer are coupled through the intra-layer couplings $\sigma$. Gaussian white noise is applied to both layers, only the noise applied to layer 1 is supra-threshold.}
\label{fig:esquema_two}
\end{figure}

We study a two-layer multiplex network schematically represented in Fig.~\ref{fig:esquema_two}. Each layer is a ring of $N$ FitzHugh-Nagumo (FHN) neurons~\cite{FHN1,FHN2} in the excitable regime with Gaussian white noise. In each layer each neuron has two neighbors, one in each direction of the ring. All the links (intra-layer and inter-layer) are diffusive and bidirectional. The model equations are:
\begin{align}
\epsilon\frac{du_{1i}}{dt}& =u_{1i}-(u_{1i})^3/3-v_{1i}+\frac{\sigma}{2}\sum_{j=i-1}^{i+1}(u_{1j}-u_{1i})+\mu_i\sigma_{12}(u_{2i}-u_{1i}) +\sqrt{2D_1}\zeta_{1i}(t),  \nonumber \\
\frac{dv_{1i}}{dt}& =u_{1i}+a, \nonumber \\
\epsilon\frac{du_{2i}}{dt}& =u_{2i}-(u_{2i})^3/3-v_{2i}+\frac{\sigma}{2}\sum_{j=i-1}^{i+1}(u_{2j}-u_{2i})+\mu_i\sigma_{12}(u_{1i}-u_{2i}) +\sqrt{2D_2}\zeta_{2i}(t), \nonumber \\
\frac{dv_{2i}}{dt}& =u_{2i}+a. \nonumber
\label{eq:fnModell}
\end{align}
Here $u_{ki}$ and $v_{ki}$ are the activator variable (i.e., voltage-like variable) and the inhibitor variable respectively; index $i$ ($i=1\dots N$) denotes the $i$-th neuron in each of the two layers while index $k$ ($k = 1,2$) denotes the layer in which the neuron is located.

The parameter $\sigma$ denotes the coupling strength between neurons in the same layer that we refer to as intra-layer coupling.  The strength of the coupling between the layers (that we refer to as inter-layer coupling)  is characterized by the parameter $\sigma_{12}$. 
Here we focus on the ``weak multiplexing'' situation, in which the inter-layer coupling is much weaker than the intra-layer coupling (i.e., $\sigma_{12} << \sigma$).
 
The deterministic bifurcation parameter is $a$. The uncoupled ($\sigma = \sigma_{12} = 0$) and deterministic ($D_1=D_2 = 0$) neurons undergo a Hopf bifurcation at $a = 1$: for $|a| < 1$ the neurons fire periodically while for $|a| > 1$ they are excitable. In this study, we focus on the situation in which all neurons are excitable and keep $a = 1.05$ and $\epsilon = 0.01$ constant.
The small parameter $\epsilon$ is responsible for the time scale separation of fast activator and slow inhibitor.

$\zeta_{1i}(t)$ and $\zeta_{2i}(t)$ represent uncorrelated Gaussian white noise sources whereas $D_1$ and $D_2$ represent the noise intensities, respectively. 
As we are interested in understanding how the activity of the neurons in layer 1 excite the neurons in layer 2, $D_1$ and $D_2$ are chosen such that $D_1$ is supra-threshold (i.e., noise induces spiking of the neurons in layer 1) while $D_2$ is sub-threshold  (i.e., neurons in layer 2 are excited through the multiplex coupling $\sigma_{12}$: if $\sigma_{12} = 0$, neurons in layer 2 do not fire).  We vary $D_1$ as a control parameter and keep $D_2 = 2.5\cdot 10^{-6}$ constant.

\section{Methods} \label{sec:methods}
To quantify coherence resonance (i.e., noise-induced regularity of the spiking activity) we use the coefficient of variation, $R$, of the distribution of inter-spike-intervals~\cite{PIK97}. It is computed for neurons in layer $k=1$ or in layer $k=2$ as

\begin{equation}
R_k=\sigma^{ISI}_k/\left\langle ISI\right\rangle_k
\end{equation}
where the mean, $\left\langle ISI\right\rangle_k$, and the standard deviation, $\sigma^{ISI}_k$, of the inter-spike intervals (ISIs) are calculated by averaging over time and over space (i.e., by averaging the inter-spike intervals in all the spike sequences of all the neurons in layer $k$). If layer $k$ shows coherence resonance, there will be a pronounced minimum of $R_k$ with respect to the noise strength $D_1$ ($D_2$ is kept fixed below the firing threshold). On the other hand, if a layer shows anti-coherence resonance, there will be a maximum of $R_k$ with respect to $D_1$~\cite{anticoh}. 

\section{Results} \label{sec:results}

We begin by analyzing the most simple configuration: one neuron in each layer (i.e., two diffusely and bidirectionally coupled FHN neurons, one with supra-threshold noise, and the other, with sub-threshold noise).

\begin{figure}[t] 
\centering
\includegraphics[width=0.15\columnwidth]{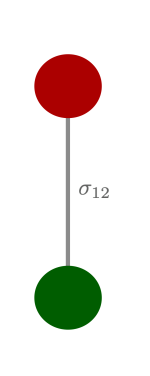}
\includegraphics[width=0.7\columnwidth]{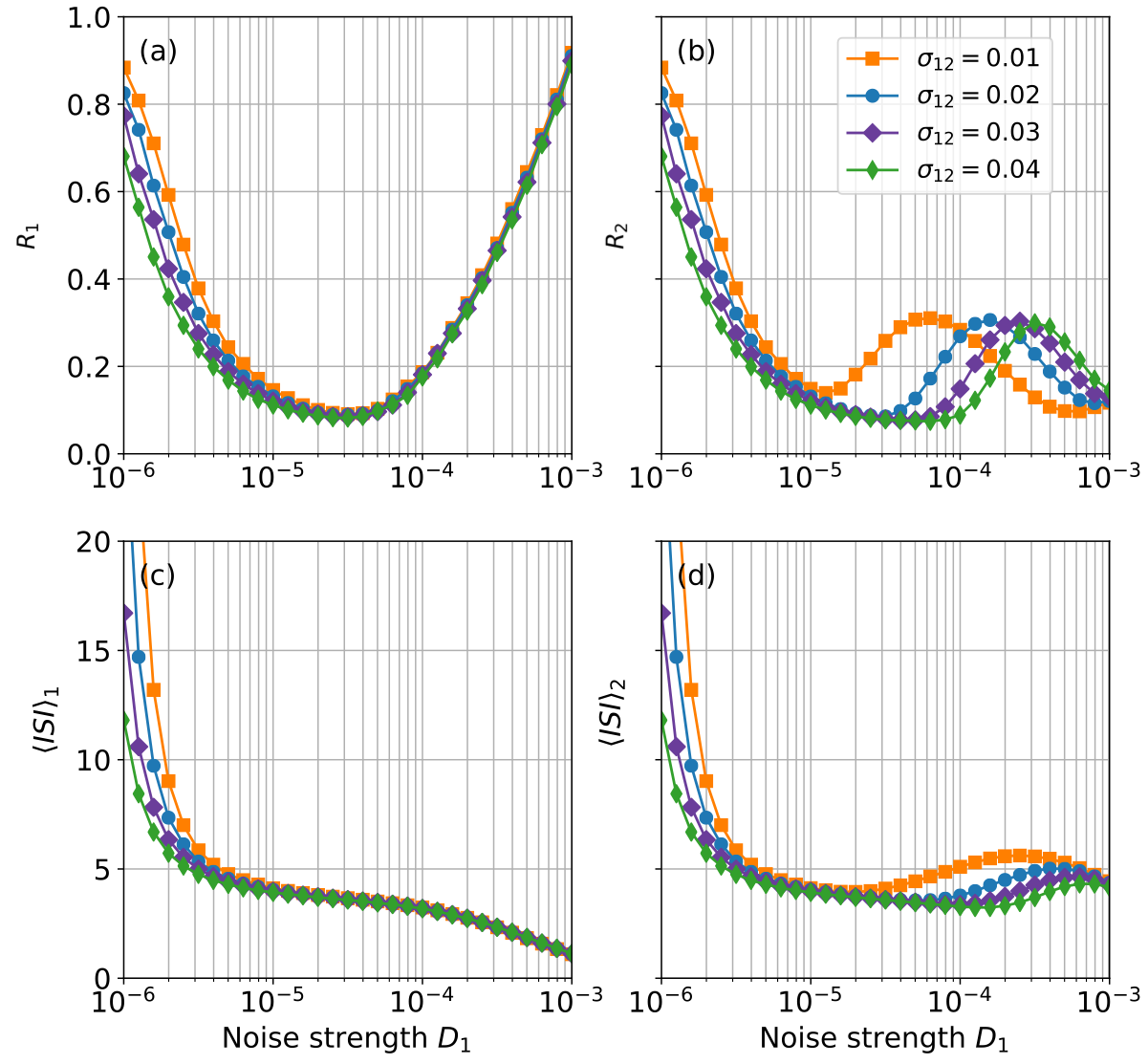}
 \caption{Characterization of the spiking activity of two coupled neurons for various values of the coupling strength. (a) $R_1$, (b) $R_2$, (c) $\left\langle ISI\right\rangle_1$ and (d) $\left\langle ISI\right\rangle_2$ as a function of the supra-threshold noise intensity of neuron 1, $D_1$. }
\label{fig:one}
\end{figure}

Figures~\ref{fig:one}(a), (b) display $R_1$ and $R_2$ vs. the level of supra-threshold noise, $D_1$, for different values of the coupling strength, $\sigma_{12}$. For neuron 1, $R_1$ shows the characteristic minimum of coherence resonance, and we see that $R_1$ is either unaffected (for strong noise) or only slightly affected (for weak noise) by the coupling to neuron 2. This is due to the fact that we consider ``weak multiplexing'', i.e., the two  neurons are weakly coupled. 

For neuron 2, $R_2$, in addition of showing the characteristic minimum of coherence resonance, displays a maximum at a higher noise level that indicates anti-coherence resonance. We also note that the coupling strength $\sigma_{12}$ affects the level of noise for which coherence and anti-coherence resonances occur: for increasing $\sigma_{12}$, both the minimum and the maximum shift to the right, i.e., towards higher noise intensity.  

In Fig.~\ref{fig:one}(c) we see that the average ISI of neuron 1 monotonically decreases with the level of noise: as expected, the supra-threshold noise induces spikes and the spiking rate increases (i.e., the average ISI decreases) with $D_1$. However, in Fig.~\ref{fig:one}(d) we see that the average ISI of neuron 2 has a non-monotonic variation with $D_1$: for high levels of noise, inverse stochastic resonance (ISR)~\cite{isr} occurs. ISR is the phenomenon by which noise inhibits neuronal activity: the spike rate is minimum (and therefore, the average ISI is maximum) at a certain level of noise.

\begin{figure}[t] 
\centering
  \includegraphics[width=0.2\columnwidth]{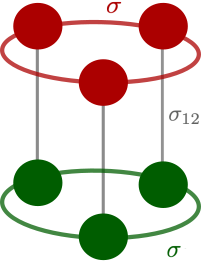}
  \includegraphics[width=0.7\columnwidth]{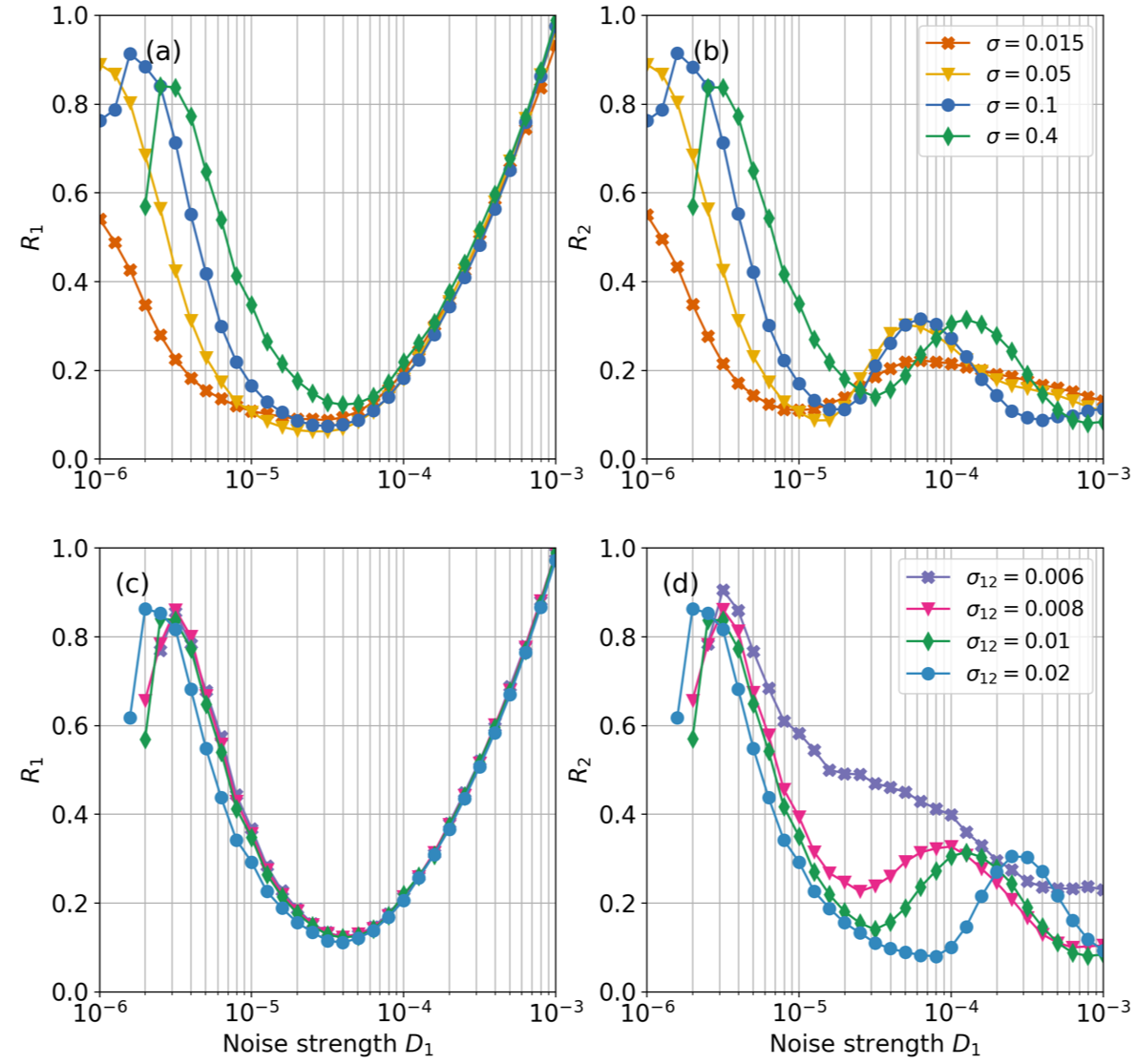}
  \caption{$R_1$ and $R_2$ as a function of the noise intensity $D_1$ for two coupled layers with $N = 3$ neurons each. In (a), (b) $\sigma_{12} = 0.01$ is kept constant and $\sigma$ is varied; in (c), (d), $\sigma = 0.4$ is kept constant and $\sigma_{12}$ is varied; other parameters are indicated in the text.}
\label{fig:two}
\end{figure}

A similar behavior is seen in Fig.~\ref{fig:two}, where we analyze two rings with $N=3$ neurons each. In layer 1, the intra-layer coupling only affects the neuronal activity when the noise is weak; for strong noise, $R_1$ is unaffected by $\sigma$ [Fig.~\ref{fig:two}(a)].
In other words, the complex interplay of coupling and noise defines the dynamics: for weak noise, the coupling dominates the dynamics and for large values of noise intensity, noise governs the dynamics. In more detail, for small noise, weak intra-layer coupling supports oscillations with higher regularity. For stronger coupling between the neurons inside the layer, it becomes harder to bring the nodes across the threshold by weak noise (i.e., the coupling dominates). For strong noise, the dynamics is governed predominantly by stochastic input and the inter-layer coupling does not have an impact on the regularity of the noise-induced oscillations.

On the other hand, $\sigma_{12}$ has almost no effect on the activity of layer 1, regardless of the noise level [Fig.~\ref{fig:two}(c)]. This is again due to the fact that the chosen coupling parameters correspond to weak multiplexing.  In layer 2 [Figs.~\ref{fig:two}(b), (d)], by tuning $\sigma$ or $\sigma_{12}$ we can achieve anti-coherence resonance for both weak and strong noise. For intermediate noise levels, if $\sigma$ or $\sigma_{12}$  are large enough, we observe coherence resonance.

\begin{figure}[t] 
\centering
\includegraphics[width=1.0\columnwidth]{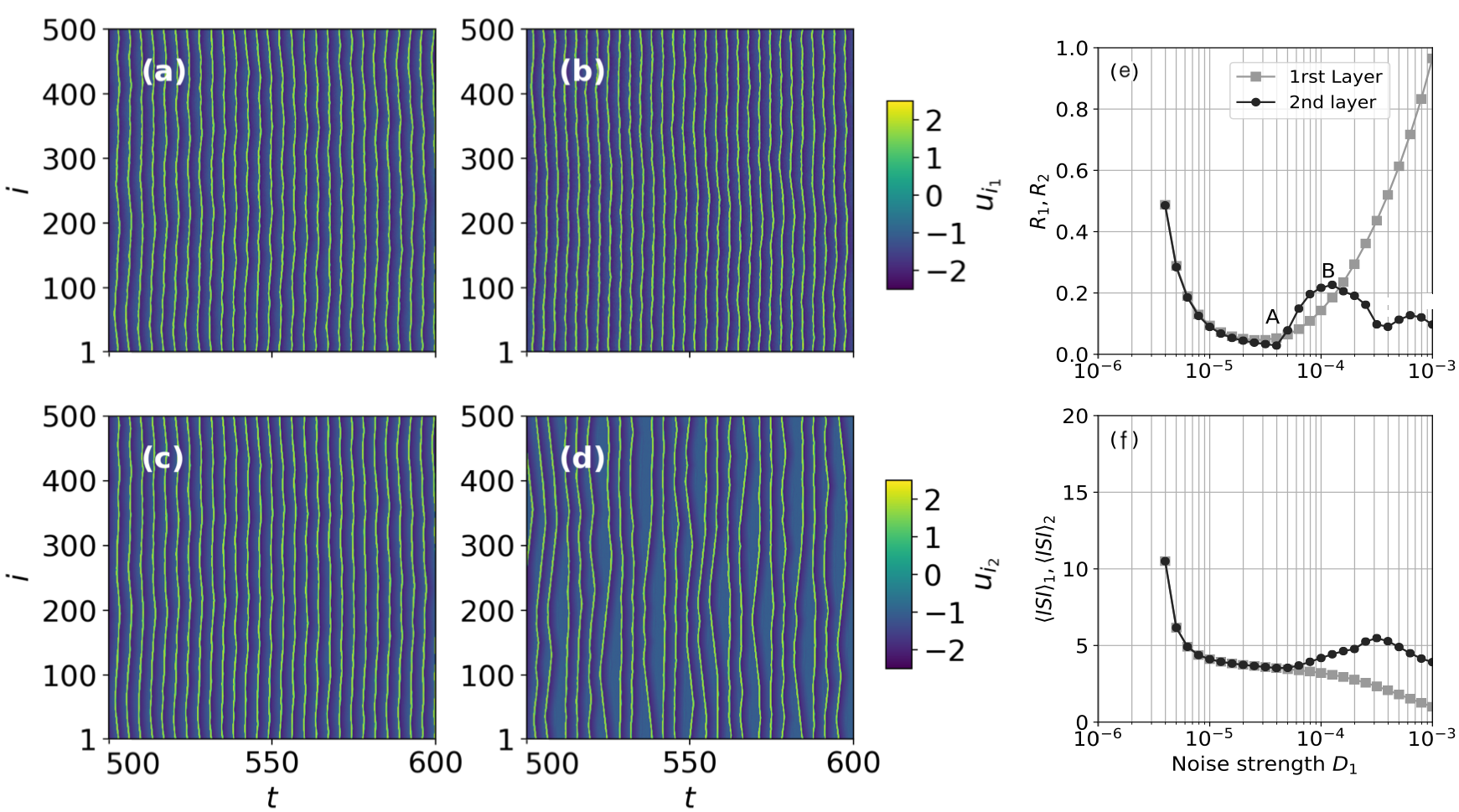}
\caption{Characterization of the spiking activity of two coupled layers with $N = 500$ neurons each. Space-time plots of neurons in layer 1 (a, b) and in layer 2 (c, d) when the noise level produces maximum (a, c) and minimum (b, d) coherence (see points labeled A and B in panel (e)). $R_1$ and $R_2$ (e) and $\left\langle ISI\right\rangle_1$ and $\left\langle ISI\right\rangle_2$ (f) as a function of the noise intensity $D_1$. The coupling strengths are $\sigma = 0.4$ and $\sigma_{12} = 0.01$, other parameters are as indicated in the text.}
\label{fig:500neurons}
\end{figure}

Next, we study the dynamics of two large inter-connected layers. We consider $N = 500$ neurons in each ring; qualitatively similar results were found for other values of $N$. 
Figure~\ref{fig:500neurons} shows that weak multiplexing induces coherence resonance in layer 2 (as shown in \cite{SEM18}).  Interestingly, the second layer presents {anti-coherence resonance} ($R_2$ displays a maximum). 
 
In order to visualize the underlying dynamics we display the activity of the neuronal populations in layer 1 and in layer 2 using space-time plots. The results are presented in Fig.~\ref{fig:500neurons}. We consider the noise levels that produce maximum or minimum regularity in layer 2 [points marked A and B, respectively in Fig.~\ref{fig:500neurons}(e)]. We see that in point A both layers have the same firing rate, and the neurons fire synchronously; in point B, the firing dynamics in layer 1 is still quite regular, while in layer 2 it is quite irregular, in space and in time. 

\begin{figure}[t] 
\centering
\includegraphics[width=0.9\columnwidth]{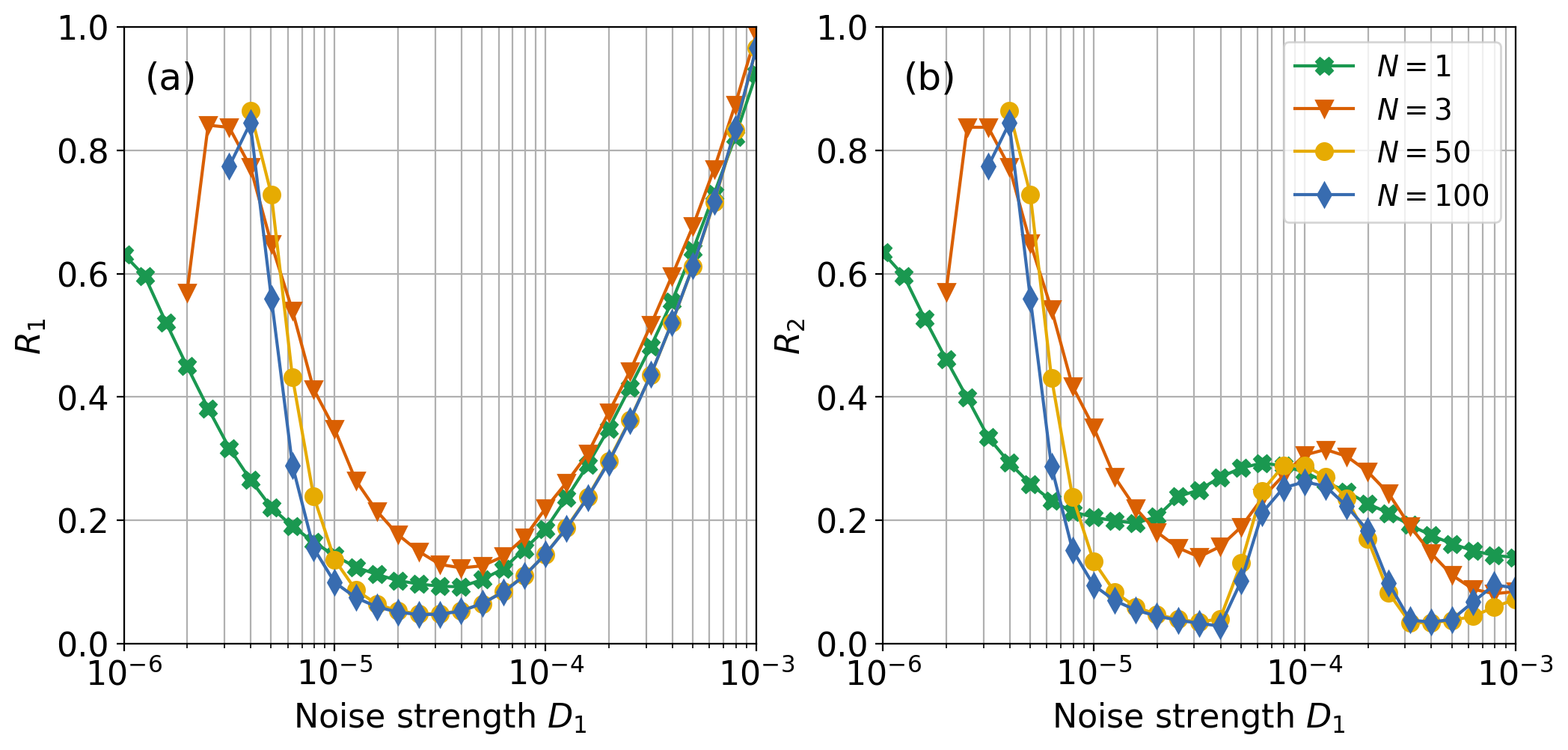}
\caption{Effect of the number of neurons, $N$, in each ring.  The coupling strengths are $\sigma = 0.4$ and $\sigma_{12} = 0.01$, other parameters are as indicated in the text.}
\label{fig:effect_N}
\end{figure}

Similar results were obtained with other network sizes.  In fact, Fig.~\ref{fig:effect_N} shows that the dynamics becomes insensitive to the number of neurons if the ring is large enough: the variation of $R_1$ and $R_2$ with the noise level is very similar for $N=50$ and $N=100$.

Figure 4(f) demonstrates that weak multiplexing induces ISR in the second layer also for larger system size ($N=500$).

\begin{figure}[t] 
\centering
\includegraphics[width=0.4\columnwidth]{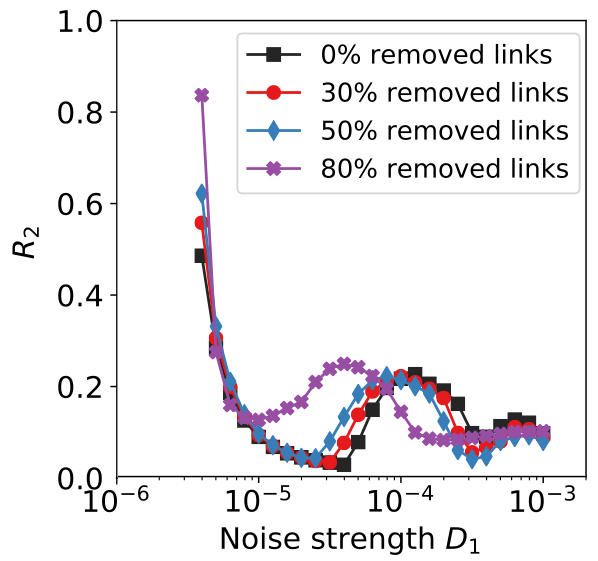}
\caption{$R_2$ as a function of the noise intensity $D_1$, when a given percentage of randomly selected inter-layer links are removed. The parameters are as in Fig.~\ref{fig:500neurons}. The variation of $R_1$ is not shown because the removal of the weak inter-layer links has almost no effect in the activity of layer 1 [the plot of $R_1$ vs. $D_1$ is very similar to that shown in Figs. 3(c), 4(e) or 5(a)]. }
\label{fig:remove}
\end{figure}

To gain further insight into the role of the weak multiplexing, and how the spiking activity of layer 1 generates a spiking activity in layer 2, we analyze the effect of randomly removing a certain percentage of inter-layer links. In Fig.~\ref{fig:remove} we see that coherence and anti-coherence resonances are induced in layer 2 even when up to 80\% of the inter-layer links are removed. The minimum amount of links that can be removed depends on the size of the rings. For example, for two rings with 50 neurons each, we could remove up to $70\%$ of the inter-layer links, and still be able to observe coherence resonance in layer 2 (not shown). 

Because we consider weak multiplexing ($\sigma_{12}<<\sigma$) link removal has almost no effect in the spiking activity in layer 1. As it was shown in Fig.~\ref{fig:two}(c) for $N=3$, $R_1$ is almost unaffected by $\sigma_{12}$, and this holds also for larger $N$. 

However, when there are only few inter-layer links, their position in the ring strongly affects the spiking activity of layer 2. For instance, three  inter-layer links in neighboring neurons can be enough to induce a spiking activity in layer 2, but the same number of inter-layer links distributed among non-neighboring neurons might not be sufficient to induce a spiking activity in layer 2. A detailed study of this effect is left for future work.

\section{Conclusions} \label{sec:conclusions}

We have studied the dynamics of two neuronal populations weakly coupled in a multiplexed configuration, and subject to different levels of noise (one population has supra-threshold noise, while the other, sub-threshold noise). The activity of the neurons was simulated with the FHN model. We found that coherence, anti-coherence and inverse stochastic resonances can be induce in layer 2 (with subthreshold noise), for appropriate  levels of supra-threshold noise in layer 1. The results were found to be robust to the number of neurons in each neuronal population.

While the coupling topology considered here is not biologically realistic, it is a simple toy model to characterize how noise-induced spiking activity in one layer can propagate and induce spiking activity in another layer. We have found that a small percentage of randomly distributed inter-layer links can be sufficient to induce spikes in the ``silent'' layer. Further work will aim at using more advanced data analysis tools, such as symbolic ordinal analysis~\cite{bp,maria,roberto}, to further characterize the regularity of the neuronal activity induced in layer 2.

Our work yields light into the complex nonlinear dynamics of excitable stochastic units coupled in a simple bilayered structure. Further work using more complex structures is of course necessary in order to advance the understanding of the role of noise and multiplexing in biologically realistic neuronal models, such as cortical networks~\cite{belen}.
 
\section*{Acknowledgments}

C.M. acknowledges partial support from Spanish Ministerio de Ciencia, Innovación y Universidades grant PGC2018-099443-B-I00 and ICREA ACADEMIA, Generalitat de Catalunya. M.M. and A.Z. acknowledge support by the Deutsche Forschungsgemeinschaft (DFG, German Research Foundation) - Project No. 163436311 - SFB 910.

\end{document}